\documentclass[prl,preprint,superscriptaddress]{revtex4}
\usepackage{color}
\usepackage{graphicx}
\usepackage{dcolumn}
\usepackage{bm}

\begin{document}


\title{Strong coupling between magnetic and structural order parameters in SrFe$_2$As$_2$}

\author{A.\,Jesche}
\author{N.\,Caroca-Canales}
\author{H.\,Rosner}
\author{H.\,Borrmann}
\author{A.\,Ormeci }
\author{D.\,Kasinathan}
\author{K.\,Kaneko}
\affiliation{Max Planck Institute for Chemical Physics of Solids, D-01187 Dresden, Germany}
\author{H.\,H.\,Klauss}
\affiliation{IFP, TU-Dresden, D-01069 Dresden, Germany}
\author{H.\,Luetkens}
\author{R.\,Khasanov}
\author{A.\,Amato}
\affiliation{Laboratory for Muon-Spin Spectroscopy, Paul-Scherrer-Institute, CH-5232 Villigen PSI, Switzerland}
\author{A.\,Hoser}
\affiliation{Hahn-Meitner-Institute, D-14109 Berlin, Germany}
\author{C.\,Krellner}
\author{C.\,Geibel}
\affiliation{Max Planck Institute for Chemical Physics of Solids, D-01187 Dresden, Germany}
\email{geibel@cpfs.mpg.de}

\date{\today}

\begin{abstract}
X-ray and Neutron diffraction as well as muon spin relaxation and M\"ossbauer experiments performed on SrFe$_2$As$_2$ polycrystalls confirm a sharp first order transition at $T_0 = 205$\,K corresponding to an orthorhombic phase distortion and to a columnar antiferromagnetic Fe ordering with a propagation vector (1,0,1), and a larger distortion and larger size of the ordered moment than reported for BaFe$_2$As$_2$. The structural and the magnetic order parameters present an remarkable similarity in their temperature dependence from $T_0$ down to low temperatures, showing that both phenomena are intimately connected. Accordingly, the size of the ordered Fe moments scale with the lattice distortion when going from SrFe$_2$As$_2$ to BaFe$_2$As$_2$. Full-potential band structure calculations confirm that the columnar magnetic order and the orthorhombic lattice distortion are intrinsically tied to each other.
\end{abstract}

                             

\maketitle

Compounds with FeAs layers have recently attracted considerable interest, because they present an intriguing magnetic and structural transition, which gets suppressed upon doping resulting in the appearance of high temperature superconductivity.  This behaviour was first observed in the RFeAsO series of compounds (R= La-Gd)\,\cite{Kamihara:2008,ChenNature:2008, delacruz:2008, Klauss:2008, Fratini:2008} and more recently in the AFe$_2$As$_2$ class of materials (A = Ba, Sr) \cite{RotterA:2008, RotterB:2008, KrellnerPRL:2008, Chen122:2008, Sasmal:2008}. The onset of superconductivity at the disappearance of a magnetic ordered state is reminiscent of the behavior in the cuprates and in the heavy fermion systems, and therefore suggests the SC state in these doped layered FeAs systems to be of unconventional nature, too. While this has to be confirmed by further studies, there seems to be a general belief that the intriguing properties of these compounds are connected with very peculiar properties of the FeAs layers.
\\
While the occurrence of magnetic order in LaFeAsO and in the AFe$_2$As$_2$ compounds has been reported already more than 15 years ago \cite{Pfisterer:1983, Raffius:1993}, the observation of the lattice deformation is quite recent \cite{delacruz:2008, Fratini:2008, RotterA:2008, Yan:2008}. The interaction between both phenomena is a very interesting problem on its own. A thorough understanding of these two phenomena, their mutual relation, and how they get suppressed under doping is likely a prerequisite to get a deeper insight into the origin and the nature of the superconducting state. In the RFeAsO compounds, the formation of the spin density wave (SDW) seems to occur in a second order transition at a slightly lower temperature $T_N \approx 140$\,K than the structural transition at $T_0=150$\,K \cite{delacruz:2008, Klauss:2008}. For BaFe$_2$As$_2$, the first report by M.\,Rotter \textit{et al.} \cite{RotterA:2008} suggested both ordering phenomena to occur simultaneously at a second order transition at $T_0=140$\,K. Shortly later, J. Q. Huang \textit{et al.} \cite{Huang:2008}, claimed the structural distortion to be first order while the magnetic order sets in continuously once the structural distortion is completed. Thus the present picture for both the RFeAsO and BaFe$_2$As$_2$ systems suggests that the structural distortion has to be completed before the AF order can form, and that the two order parameters are not directly connected. For SrFe$_2$As$_2$, we recently showed that a high quality sample presents a very sharp first order transition at $T_0=205$\,K, without any evidence for a second transition \cite{KrellnerPRL:2008}. In the present, paper we report a precise study of the evolution of the magnetic and of the structural order parameter in this compound by combining temperature dependent muon spin resonance and X-ray diffraction measurements with bulk susceptibility, resistivity, specific heat as well as preliminary 
M\"ossbauer and neutron scattering data. Our results demonstrate that in SrFe$_2$As$_2$, the formation of the SDW and the lattice distortion are intimately coupled. Comparison with results reported for BaFe$_2$As$_2$ also supports a strong connection between both order parameters.
\\
The sample preparation and characterization have been described in detail in our previous paper \cite{KrellnerPRL:2008}. Susceptibility, specific heat and resistivity measurements were carried out using standard techniques in commercial equipments PPMS and MPMS of Quantum Design.
Temperature dependent X-ray powder pattern were obtained using an imaging plate Guinier Camera HUBER G670 (Co-K$_{\alpha}$ radiation) equipped with a closed cycle cryostat. 
Zero field muon spin relaxation (ZF-$\mu$SR) experiments were performed between 1.6\, and 300\,K using the GPS spectrometer at the Paul Scherrer Institute. To gain deeper insight into the relation of magnetism and the orthorhombic distortion in AFe$_2$As$_2$ on a microscopic level, we performed density functional band structure calculations within the
local (spin) density approximation (L(S)DA). Using the experimental structural parameters of the tetragonal cell \cite{Pfisterer:1983,Raffius:1993,Pfisterer:1980} as a starting point, we applied the full-potential local-orbital code FPLO \cite{koepernik:1999} (version 7.00-28) in both scalar-relativistic and fully relativistic versions, respectively, with the Perdew-Wang exchange correlation potential \cite{Perdew:1992}. A well-converged $k$-mesh of at least 18$^3$ points within the Brillouin zone of the larger orthorhombic cell has been used.
\\
In Fig. 1a we show the anomalies in the resistivity $\rho(T)$, susceptibility $\chi(T)$ and specific heat $C(T)$ which evidence a sharp, first order transition in our polycrystalline SrFe$_2$As$_2$ sample, as discussed in our previous paper \cite{KrellnerPRL:2008}.
While $\rho(T)$ is only weakly decreasing with temperatures between 300\,K and 205\,K, it presents a 5\% drop at $T_0$ followed by a further strong decrease to low temperatures, resulting in a resistivity ratio $RR_{1.8 K}\approx 32$. The susceptibility, except for a Curie like contribution likely due to paramagnetic impurities or a small amount of foreign phases, seems to be $T$ independent above and below $T_0$, but presents also a drop of $\Delta\chi\approx 1.1 \cdot 10^{-9}$\,m$^3$/mol at $T_0$. 
The specific heat measurement shows a sharp peak at $T_0$, which was interpreted as first order transition with a latent heat $\Delta H \approx 200$\,J/mol. 
We shall first focus on the results of the X-ray measurements.
At room temperature and down to 210\,K the powder diffraction pattern evidenced an undistorted tetragonal (TT) ThCr$_2$Si$_2$ structure type. In contrast, in all pattern taken at 205\,K or lower temperatures, some of the Bragg peaks are well splitted, while others are not, demonstrating the structural distortion (Fig.\,1b). The spectra at 205\,K and below can be well fitted with an orthorhombic (OT) unit cell (Fmmm) with $a_{OT}$ = $a_{TT}\cdot \sqrt{2} (1 + \delta)$ and $b_{OT}$ = $a_{TT} \cdot \sqrt{2}(1 - \delta)$ in analogy to the structure proposed for BaFe$_2$As$_2$ \cite{RotterA:2008} and in accordance with \cite{Yan:2008}. So $\delta$ corresponds to the order parameter of the structural phase transition. A lattice parameter fit at the lowest investigated temperature $T = 60$\,K gave $a = 5.5746 (4)$\,\AA, $b = 5.5130(8)$\,\AA\,and $c = 12.286(4)$\,\AA\,corresponding to a saturation value of the distortion $\delta_0 = 0.56(1) \cdot 10^{-2}$ at low $T$. The evolution of $\delta$ with temperature was determined by analyzing precisely the splitting of the 400/040 Bragg peaks (Fig.\,1b and Fig.\,2a). Here we included data taken upon cooling and data taken upon heating the sample. We did not observe any differences between both sets of data. Between 210\,K and 205\,K, the 220 peak of the TT high temperature phase disappears abruptly, being replaced by the 400 and 040 peaks of the OT low temperature phase.
At 210\,K, shoulders on both sides of the 220 peak indicate that a small amount of OT phase is coexisting with the TT phase, in accordance with a first order transition. The presence of this OT phase above $T_0$ might be due to strain or defects induced by the powdering process.
The distortion $\delta$ increases step like to 70\% of $\delta_0$. This is a further clear evidence for a first order transition. However, $\delta$ continues to increase with decreasing temperatures, indicating a further strengthening of the order parameter below the transition. A comparison with the data reported previously by J. Q. Yan \textit{et al.} \cite{Yan:2008} gives a strong evidence that this further increase of $\delta(T)$ below $T_0$ is an intrinsic property and not just a consequence of an imperfect sample. In general both sets of data are similar 
\cite{note1}.
However, our results evidence a very abrupt transition from the TT to the OT phase, while the data of \cite{Yan:2008} show a large coexistence region ranging from 160\,K up to 198\,K. This broadening of the transition as well as the lower $T_0$ in the 122 single crystals of \cite{Yan:2008} are due to Sn incorporation. However, both the absolute value of the splitting at low $T$ and that at the transition are very similar to our results. Thus, while the transition temperature and the sharpness of the transition are quite sensitive to defects, the splitting at $T_0$ and at $T = 0$\,K as well as the increase of $\delta(T)$ below $T_0$ are not.
\\
In order to address the magnetic order parameter, we performed preliminary Fe M\"ossbauer measurements and elastic neutron scattering experiments. The M\"ossbauer spectra evidenced very well defined hyperfine splitting at low $T$ corresponding to an hyperfine field of 8.5\,T \cite{Luetkens:2008}, which is identical to the value reported for EuFe$_2$As$_2$ \cite{Raffius:1993}. In the neutron scattering spectra we observed sharp magnetic Bragg peaks below $T_0$, similar to those reported for BaFe$_2$As$_2$. 
The higher precision of our measurement allowed to uniquely fix the magnetic structure as a columnar antiferromagnetic order with propagation vector (1,0,1) and Fe moment of 1.0(1) $\mu_{B}$ oriented along the a-axis.   
However, we found that the most precise information on the evolution of the magnetic order parameter was obtained in our $\mu$SR experiments. Muon spin relaxation is a well established method for revealing and studying magnetic order. It probes the local field induced at the site(s) of the muon by slowly fluctuating or ordered nearby magnetic moments. For temperatures above 205\,K we observe only a slow decay of the muon polarization as expected for a non-magnetic material. Below 205\,K, well defined and strong oscillations appear in the time dependence of the muon polarization, as shown in the inset of Fig.\,3, evidencing a precession of the muon in an internal field. A Fourier analysis of the signal reveals two distinct components with very well defined frequencies, one at $f_1 = 44$\,MHz corresponding to $\approx$\,70\% of the signal and one at $f_2 = 13$\,MHz corresponding to $\approx$\,30\% of the signal \cite{Luetkens:2008}. This indicates the presence of two distinct muon sites, one being more strongly and one more weakly coupled to the Fe moments. This resembles the situation in LaFeAsO where also two components, one with a larger frequency $f_1 = 23$\,MHz corresponding to 70\% of the muons and one with a lower frequency $f_2 = 3$\,MHz corresponding to 30\% of the muons were observed \cite{Klauss:2008}. The ratio between the respective $f_1$ frequencies in SrFe$_2$As$_2$ and LaFeAsO is similar to the ratio of the hyperfine field measured in M\"ossbauer experiments and thus to the ratio of the ordered Fe Moments. This suggests that the muon site corresponding to $f_1$ is the same in both types of compounds and likely located within the FeAs layers, while the muon site corresponding to $f_2$ is probably in the region separating the FeAs layers, which differs between both types of compounds. The oscillation we observed in SrFe$_2$As$_2$ are much better defined than those reported for LaFeAsO, which is likely related to a much better crystallinity and higher homogeneity of the AFe$_2$As$_2$ compounds compared to the RFeAsO ones. On the other hand it indicates that the internal field at each muon site in SrFe$_2$As$_2$ is sharply defined, implying a well defined long range commensurate magnetic order.
In the main part of Fig.\,3, we show the temperature dependence of $f_1$ in SrFe$_2$As$_2$. $f_1$ is proportional to the size of the ordered moment and thus to the magnetic order parameter. In contrast to LaFeAsO where $f_1$ is increasing continuously below a second order transition at $T_N \approx 134$\,K, we observe in SrFe$_2$As$_2$ at 205\,K a sharp step like increase of $f_1$ to 66\% of its saturation value at low $T$. This is again an indication for a first order transition. However, as already noticed for the $T$ dependence of the lattice distortion $\delta$, also the magnetic order parameter further increases below $T_0$ with decreasing $T$. We compare in Fig.\,2b the $T$ dependence of $\delta(T)$ and $f_1(T)$ normalized to their saturation values at low $T$. The $T$ dependencies are identical within the accuracy of the experiments. This demonstrates that both order parameters are intimately coupled to each other.
To elucidate the role of various possible magnetic orderings for the
OT distortion of the crystal structure for SrFe$_2$As$_2$
and the related Ba compound, we performed band structure calculations
for various spin configurations within the FeAs layers. Starting from
different initial ordering patterns, we obtained self consistent
solutions for (i) non-magnetic, (ii) ferromagnmetic, (iii) Ne\'el
ordered and (iv) columnar ordered FeAs layers. For both systems the
lowest energy was found for the columnar ordered state. Starting from
the experimental structural parameters for the TT unit cells
we varied the axis ratio $b/a$, keeping the other parameters and the
cell volume constant. The resulting curves for the Ne\'el ordered and
columnar ordered FeAs layers are shown in Fig.~\ref{lda}.
Except for the columnar magnetic order (iv) that yields a significant OT split for the TT axes, all other patterns (i-iii) resulted in an energy minimum for an undistorted TT structure. The inclusion of spin-orbit coupling did not change this result within the numerical error bars. In surprisingly good agreement with our neutron experiments, we obtain a shortening of the $b$ axis along the ferromagnetic columns compared to the $a$ axis along the antiferromagmetic propagation, resulting in a $b/a$ ratio of 0.984 for SrFe$_2$As$_2$ and 0.987 for the Ba system. These values are only slightly larger than the experimentally observed distortions
(extrapolated to zero temperature) and in excellent agreement with respect to the relative changes between both compounds. Thus, obtaining an OT axes split for the columnar magnetic order only, together with its lowest energy indicates that this magnetic order and the OT lattice distortion in both compounds are intrinsically  tied to each other.

In summary, we report a detailed study of the structural distortion and of the magnetic ordering using XR diffraction and $\mu$SR experiments as well as preliminary neutron scattering and M\"ossbauer spectroscopy data. We confirm the low temperature phase to be analogous to that reported for BaFe$_2$As$_2$ with an OT structural distortion, space group Fmmm, and a columnar antiferromagnetic ordering of the Fe moment with a propagation vector (1,0,1). 
However, both the structural distortion and the size of the ordered Fe moment are larger in the Sr than in the Ba compound. The magnetic and the structural order parameters do not only show a sharp first order transition at $T_0$ as previously suggested, but evidence the same $T$ dependence in the whole $T$ range from $T_0$ down to lowest temperatures. At $T_0$ both the OT distortion $\delta$ and the muon precession frequency $f_1$ jump to only $\approx$\,68\% of their low $T$ saturation value. A comparison with X-ray data obtained on singe crystals with a lower $T_0$ and a broader transition indicates that the further increase of $\delta(T)$ and $f_1(T)$ below $T_0$ is an intrinsic behavior and not due to defects. The identical $T$ dependence of $\delta(T)$ and $f_1(T)$ proves that the structural and the magnetic order parameters are intimately coupled.
In this respect, our data unambiguously indicate that SrFe$_2$As$_2$ behaves very differently from the picture presently proposed for the RFeAsO compounds, where the SDW is suggested to form in a second order transition at $\approx10$\,K below the structural transition, the two order parameters being disconnected. However, at the present level of investigations, one cannot exclude the double, second order type transitions in RFeAsO to be one broadened first order transition due to a poorer sample quality. A better quality of the SrFe$_2$As$_2$ sample is evidenced by the much higher residual resistivity ratio and the sharpness of the transition in all investigated properties. One of the reasons for this better quality is that the preparation of the RFeAsO compounds and especially the control of their stoichiometry is more difficult than that of the AFe$_2$As$_2$ compounds.
The strong connection between the magnetic and the structural parameter is not only present in SrFe$_2$As$_2$, but seems to be a more general property of the AFe$_2$As$_2$ systems. This is evidenced by a comparison of the magnitude of both order parameters between SrFe$_2$As$_2$ and BaFe$_2$As$_2$. From the data of Rotter \textit{et al.} one can deduced $\delta_0 = 0.36 \cdot 10^{-2}$ for BaFe$_2$As$_2$ \cite{RotterA:2008}, which is 37\% smaller than $\delta_0 = 0.56 \cdot 10^{-2}$ in SrFe$_2$As$_2$. The value of the hyperfine field determined in Fe M\"ossbauer experiments, and thus the size of the ordered Fe moment, also decreases by 36\% from $B_{eff} = 8.5$\,T in SrFe$_2$As$_2$ to $B_{eff} = 5.4$\,T in BaFe$_2$As$_2$[6]. Thus, both the magnetic and the structural order parameters scale by about the same amount when going from SrFe$_2$As$_2$ to BaFe$_2$As$_2$. Fully-relativistic band structure calculations obtain an OT lattice distortion for the columnar magnetic order, only, in very good
agreement with the experimental data, including the correct
orientation of the Fe moments along the $a$-axis. This yields strong
support to the idea that lattice distortion and the columnar magnetic order in
these compounds are intrinsically tied to each other.
While finalizing our paper, a study of the structural distortion in SrFe$_2$As$_2$ and EuFe$_2$As$_2$ appeared as a preprint, showing similar structural data but suggesting a second order type transition \cite{Tegel:2008}.

\newpage
\begin{figure}[t]
\includegraphics[width=8cm]{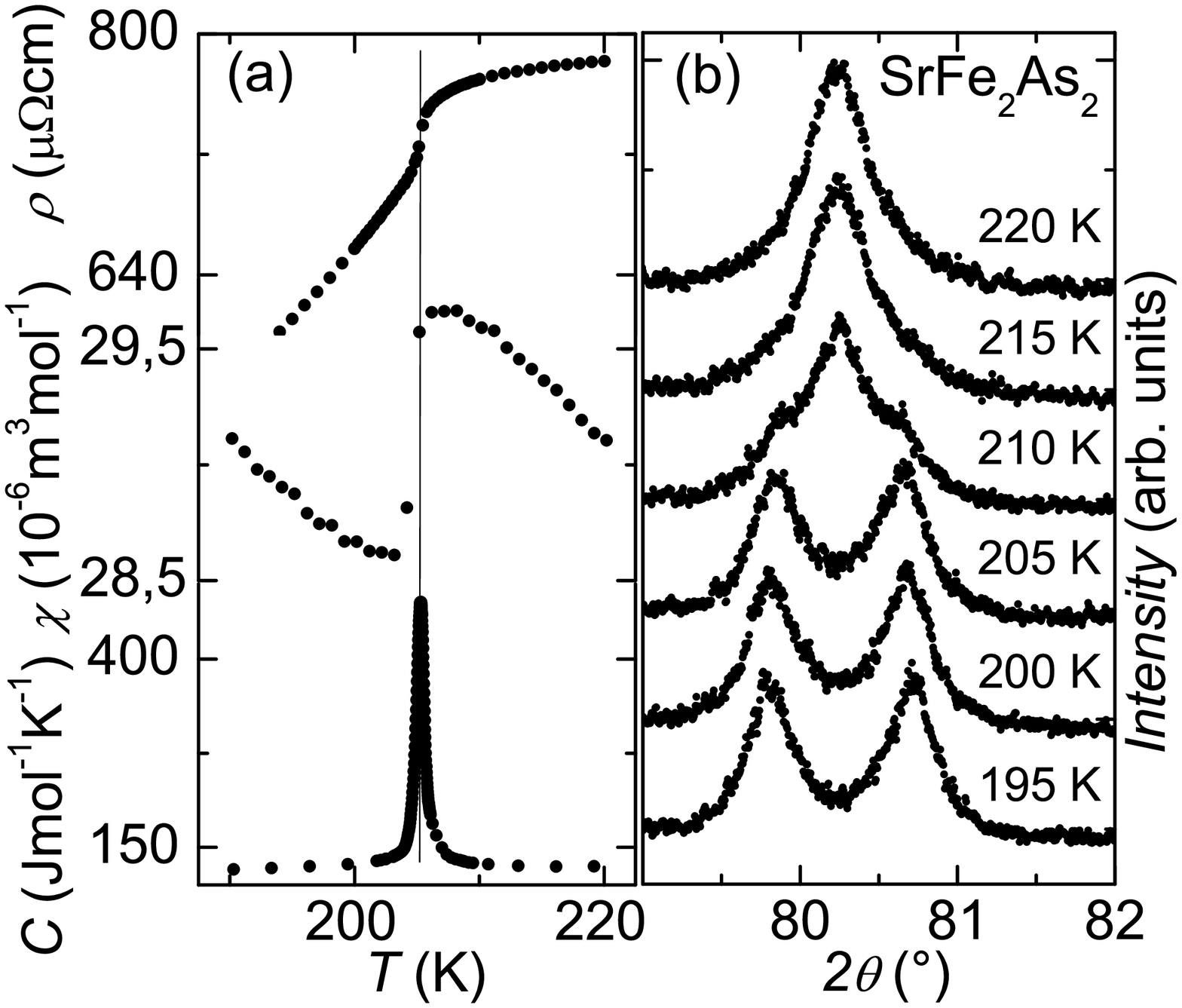}
 \caption{\label{rhochiCpeak} Left panel: resistivity $\rho(T)$, susceptibility $\chi(T)$ and  specific heat $C(T)$ in SrFe$_2$As$_2$ near the first order transition at $T_0 = 205$\,K. Right panel: splitting of the 220 tetragonal peak into the 400 and 040 peaks of the orthorhombic structure below $T_0$.}
\end{figure}

\begin{figure}[t]
\includegraphics[width=8cm]{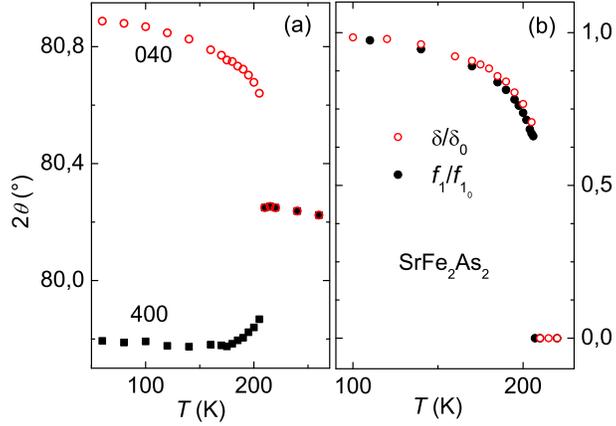}
  \caption{\label{ordnungsparameter} (Color online) Left panel: $T$ dependence of the positions of the 400 and 400 peaks. Right panel: $T$ dependence of the lattice distortion $\delta(T)$ and of the muon precession frequency $f_1$ normalized to their saturation value at low $T$. }
\end{figure}

\begin{figure}[t]
\includegraphics[width=8cm]{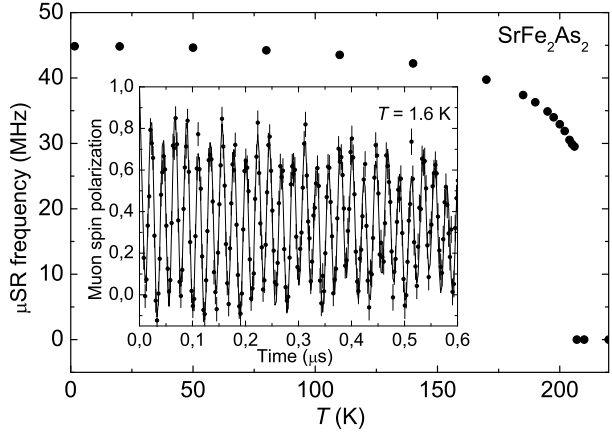}
  \caption{\label{muSR} Temperature dependence of the muon precession frequency $f_1$. Inset: Time dependence of the muon spin polarization at $T = 1.6$\,K.}
\end{figure}

\begin{figure}[t]
\includegraphics[height=8.5cm,angle=270]{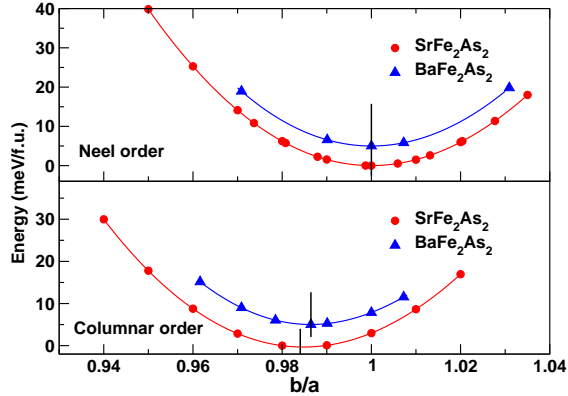}
  \caption{\label{lda} (Color online) 
Calculated total energy versus axes ratio $b/a$ for the orthorhombic
unit cell of SrFe$_2$As$_2$ (red) and BaFe$_2$As$_2$ (blue). The
calculated data points are marked by the symbols, the lines are fourth
order polynomial fits. The minima are marked by vertical lines. The minimum of SrFe$_2$As$_2$ is chosen as energy zero and the BaFe$_2$As$_2$ curves are shifted upwards by 5\,meV.
The upper panel shows no distortion for a Ne\'el order within the FeAs layers, whereas the
lower panel demonstrates the orthorhombic distortion for columnar order
in the FeAs layer.}
\end{figure}
\end{document}